\documentclass[conference]{IEEEtran}
\IEEEoverridecommandlockouts

\usepackage{cite}
\usepackage{amsmath,amssymb,amsfonts}
\usepackage{balance}
\usepackage[ruled,linesnumbered,vlined]{algorithm2e}
\usepackage{graphicx}
\usepackage{textcomp}

\usepackage{xcolor}
\usepackage{colortbl}  
\usepackage{booktabs}  
\usepackage{caption}   
\usepackage{subcaption} 
\definecolor{nvgreen}{RGB}{118,185,0}
\definecolor{amdred}{RGB}{237,28,36}
\definecolor{intelblue}{RGB}{0,113,197}
\usepackage{url}       
\usepackage{tabularx}  
\usepackage{enumitem}  
\usepackage{listings}  
\usepackage{multirow}  
\usepackage{pifont}    
\usepackage{makecell}  
\usepackage{pgfplots}  
\pgfplotsset{compat=1.18}

\DeclareFixedFont{\ttm}{T1}{txtt}{m}{n}{9}   
\DeclareFixedFont{\ttb}{T1}{txtt}{bx}{n}{9}  

\definecolor{verylightgray}{rgb}{0.95,0.95,0.95}
\definecolor{codekw}{rgb}{0.0,0.0,0.55}        
\definecolor{codecomment}{rgb}{0.25,0.50,0.35}  
\definecolor{codestr}{rgb}{0.58,0.0,0.82}       
\definecolor{codeannot}{rgb}{0.7,0.13,0.13}     
\definecolor{codefile}{rgb}{0.12,0.47,0.71}     
\definecolor{darkgreen}{rgb}{0.0,0.5,0.0}       
\definecolor{stallred}{RGB}{200,40,40}           
\definecolor{rajacolor}{RGB}{140,80,170}         
\definecolor{usercolor}{RGB}{30,100,60}          
\definecolor{annotgray}{RGB}{100,100,100}        
\usepackage{microtype}  
\usepackage{dblfloatfix} 
\usepackage{tikz}      
\usepackage{adjustbox} 
\usetikzlibrary{shapes,arrows.meta,positioning,calc,backgrounds,fit}
\usepackage{hyperref}
\hypersetup{colorlinks=false, pdfborder={0 0 0}}
\usepackage[noabbrev,capitalise]{cleveref}
\usepackage{siunitx}
\DeclareSIUnit\TFLOPS{TFLOPS}
\DeclareSIUnit\byte{B}

\usepackage{xspace}
\newcommand{\instr}[1]{\texttt{#1}}       
\newcommand{\file}[1]{\texttt{#1}}        
\newcommand{\code}[1]{\texttt{#1}}        
\newcommand{\metric}[1]{\texttt{#1}}      
\newcommand{\tool}[1]{\texttt{#1}}        

\definecolor{SysBlue}{rgb}{0.12156862745098039,0.47058823529411764,0.7058823529411765}      
\definecolor{SysBlueLt}{rgb}{0.6509803921568628,0.807843137254902,0.8901960784313725}       
\definecolor{SysOrange}{rgb}{1.0,0.4980392156862745,0.0}                                     
\definecolor{SysOrangeLt}{rgb}{0.9921568627450981,0.7490196078431373,0.43529411764705883}   
\definecolor{SysGray}{RGB}{100,100,100}                                                      
\definecolor{SysGrayLt}{RGB}{217,217,217}                                                    
\definecolor{SysYellowLt}{rgb}{1.0,1.0,0.6}                                                  
\definecolor{SysPurple}{rgb}{0.41568627450980394,0.23921568627450981,0.6039215686274509}    
\definecolor{SysTeal}{rgb}{0.2,0.6274509803921569,0.17254901960784313}                       
\definecolor{SysRed}{rgb}{0.8901960784313725,0.10196078431372549,0.10980392156862745}       
\definecolor{SysRedLt}{rgb}{0.984313725490196,0.6039215686274509,0.6}                        
\definecolor{SysGreen}{rgb}{0.2,0.6274509803921569,0.17254901960784313}                      
\definecolor{SysGreenLt}{rgb}{0.6980392156862745,0.8745098039215686,0.5411764705882353}     
\definecolor{SysViolet}{rgb}{0.41568627450980394,0.23921568627450981,0.6039215686274509}    
\definecolor{SysVioletLt}{rgb}{0.792156862745098,0.6980392156862745,0.8392156862745098}     
\definecolor{SysBrown}{rgb}{0.6941176470588235,0.34901960784313724,0.1568627450980392}      

\newif\ifblind
\blindfalse  

\begin{document}

\title{LEO: Tracing GPU Stall Root Causes via Cross-Vendor Backward Slicing}

\ifblind
  \author{\IEEEauthorblockN{Anonymous Author(s)}}
\else
  \author{
    \IEEEauthorblockN{Yuning Xia, John Mellor-Crummey}
    \IEEEauthorblockA{Rice University, Houston, TX, USA\\
    \{yuning.xia, johnmc\}@rice.edu}
  }
\fi

\maketitle

\begin{abstract}
More than half of the Top 500 supercomputers employ GPUs as accelerators. On GPU-accelerated platforms, developers face a key diagnostic gap: profilers show source lines where stalls occur, but not why they occur. Furthermore, the same kernel may have different stalls and underlying causes on different GPUs. This paper presents LEO, a root-cause analyzer for NVIDIA, AMD, and Intel GPUs that performs backward slicing from stalled instructions, considering dependencies arising from registers as well as vendor-specific synchronization mechanisms. LEO attributes GPU stalls to source instructions with the goal of explaining root causes of these inefficiencies. Across 21 workloads on three GPU platforms, LEO-guided optimizations deliver geometric-mean speedups of 1.73$\times$--1.82$\times$. Our case studies show that (1) the same kernel may require different optimizations for different GPU architectures, and (2) LEO’s structured diagnostics improve code optimization with large language models relative to code-only and raw-stall-count baselines.
\end{abstract}


\section{Introduction}
\label{sec:intro}

Today, more than half of the Top 500 supercomputers employ GPUs as accelerators~\cite{top500nov2025}. For example, supercomputers at US DOE laboratories typically employ AMD (Frontier and El Capitan), Intel (Aurora), or NVIDIA (Perlmutter and Polaris) GPUs as their primary compute engines.
Understanding GPU performance requires instruction-level visibility into where cycles are spent, which instructions stall, and which earlier instructions cause those stalls.
While GPU profiling APIs by AMD, Intel, and NVIDIA now support program counter (PC) sampling, current tools do not adequately address this need.

Vendor-provided GPU profilers, such as NVIDIA's Nsight Compute~\cite{nsightcompute2024}, AMD's rocprofv3~\cite{rocprofv32024}, and Intel's VTune~\cite{vtune2024}/unitrace~\cite{umar2023pti}, can report stall distributions measured using PC sampling per GPU instruction.
However, these tools show \emph{where} stalls occur but not \emph{why}: they present stall breakdowns without identifying which earlier instructions actually caused the observed stalls.
Research tools such as GPA (GPU Performance Advisor)~\cite{zhou2021gpa} pioneered backward slicing for GPUs, but GPA supports only NVIDIA GPUs and cannot trace memory access dependencies through synchronization instructions such as AMD's \instr{s\_waitcnt}.

To date, no tool has provided instruction-level root-cause analysis for GPUs from multiple vendors.
Each vendor exposes a different PC-sampling interface, stall taxonomy, and API, making instruction-level comparisons across architectures difficult (\cref{sec:background}).
As a result, when the same kernel is compute-bound on one platform but memory-bound on another, vendor-specific tools do not make that divergence easy to see.
Recent cross-platform studies confirm this gap: Davis et al.~\cite{davis2025taking} find that portable programming models such as RAJA~\cite{beckingsale2019raja} and Kokkos~\cite{edwards2014kokkos} do not guarantee performance portability~\cite{pennycook2019implications} across GPU vendors, and Kwack et al.~\cite{kwack2025ai} report that performance assessment tools on Frontier, Aurora, and Polaris frequently frustrated efforts to diagnose cross-platform bottlenecks.

Manually tracing backward from symptoms to root causes is tedious and error-prone.
In complex HPC applications, a stalled instruction (the symptom) and its antecedent (the root cause) may reside in different functions or source files. For example, an arithmetic operation in a function may stall while awaiting completion of a load issued by a vector template.
Without tool support, developers must manually trace register dependencies through disassembly, a process that is both tedious and error-prone.
Moreover, register-based tracing will not work for AMD \instr{s\_waitcnt} instructions, which await the completion of memory accesses, because they do not expose explicit register dependencies.

This paper presents LEO, a cross-vendor root-cause analyzer for GPU stalls on AMD, Intel, and NVIDIA GPUs.
LEO's contribution is not simply broader platform coverage. Rather, it combines three capabilities that prior work does not provide together: (1)~a unified instruction-level analysis across NVIDIA, AMD, and Intel PC-sampling ecosystems; (2)~synchronization-aware tracing through vendor-specific wait mechanisms, including AMD \instr{s\_waitcnt}, NVIDIA hardware barriers, and Intel software scoreboard (SWSB) tokens; and (3)~cross-vendor diagnosis that explains why the same source kernel may have different bottlenecks on different GPUs.
LEO builds dependency graphs from instruction-level dataflow, applies a four-stage pruning pipeline, and attributes blame with inverse-distance weighting.
LEO leverages performance measurements collected by HPCToolkit~\cite{adhianto2024refining} and builds on its cross-vendor PC-sampling infrastructure.

This paper makes the following contributions:

\begin{enumerate}[nosep,leftmargin=*]
  \item \textbf{Cross-vendor root-cause analysis.} We present a methodology that uses backward slicing to identify root-cause instructions from stalled instructions across AMD, Intel, and NVIDIA GPUs (\cref{sec:frontend}).

  \item \textbf{Explicit memory dependency tracing.} We extend the dependency graph with vendor-specific synchronization edges, including AMD \instr{s\_waitcnt} counters, Intel SWSB tokens, and NVIDIA hardware barrier bits (B1--B6) so analysis can identify memory accesses causing synchronization stalls (\cref{sec:backslicing-sync}).

  \item \textbf{Cross-vendor evaluation.} We evaluate the methodology across three GPU platforms (AMD MI300A, Intel PVC, NVIDIA GH200) on HPC and ML workloads.
  Even for the same code, root causes of stalls differ across architectures; nevertheless, LEO's insights enabled us to achieve geometric-mean speedups of 1.73$\times$--1.82$\times$ on the kernels we studied.
  We also evaluate how LEO's structured diagnostics can support large language model-based automated code tuning (\cref{sec:eval}).
\end{enumerate}

\newcommand{\cmark}{\ding{51}}%
\newcommand{\xmark}{\ding{55}}%
\newcommand{\pmark}{\ensuremath{\circ}}%
\newcommand{\na}{--}%
\newcommand{\gramark}{\cellcolor{SysBlueLt}}%

\begin{table*}[t]
  \centering
  \footnotesize
  \caption{Capability comparison of instruction-level GPU performance-analysis tools.
  \cmark: explicit support.
  \pmark: limited/partial support.
  \na: not reported or outside the tool's scope.
  LEO is the only tool implemented across NVIDIA, AMD, and Intel; broader cross-platform profilers such as DeepContext~\cite{zhao2025deepcontext} and PASTA~\cite{lin2026pasta} are discussed in \cref{sec:related}.}
  \begin{tabularx}{\textwidth}{l *{5}{>{\centering\arraybackslash}X}}
    \toprule
    \textbf{Tool}
      & \makecell{\textbf{Stall Site}\\\textbf{(src/ISA)}}
      & \makecell{\textbf{Direct}\\\textbf{Cause}}
      & \makecell{\textbf{Transitive}\\\textbf{Chain}}
      & \makecell{\textbf{Wait/Barrier}\\\textbf{Tracing}}
      & \makecell{\textbf{Quantitative}\\\textbf{Attribution}} \\
    \cmidrule(lr){2-6}
    \midrule
    Nsight Compute~\cite{nsightcompute2024}        & \cmark & \na    & \na    & \na    & \na    \\
    rocprofv3~\cite{rocprofv32024}                 & \cmark & \na    & \na    & \na    & \na    \\
    VTune~\cite{vtune2024}/unitrace~\cite{umar2023pti} & \cmark & \cmark & \na    & \pmark & \na    \\
    GPA~\cite{zhou2021gpa}                         & \cmark & \cmark & \na    & \cmark & \cmark \\
    GPUscout~\cite{sen2023gpuscout}                & \cmark & \na    & \na    & \na    & \na    \\
    DrGPU~\cite{hao2023drgpu}                      & \cmark & \na    & \na    & \na    & \na    \\
    \midrule
    \rowcolor{SysBlueLt}\textbf{LEO (this work)}    & \cmark & \cmark & \cmark & \cmark & \cmark \\
    \bottomrule
  \end{tabularx}
  \label{tab:capability-comparison}
\end{table*}

\begin{figure}[t]
\centering
\includegraphics[width=\columnwidth]{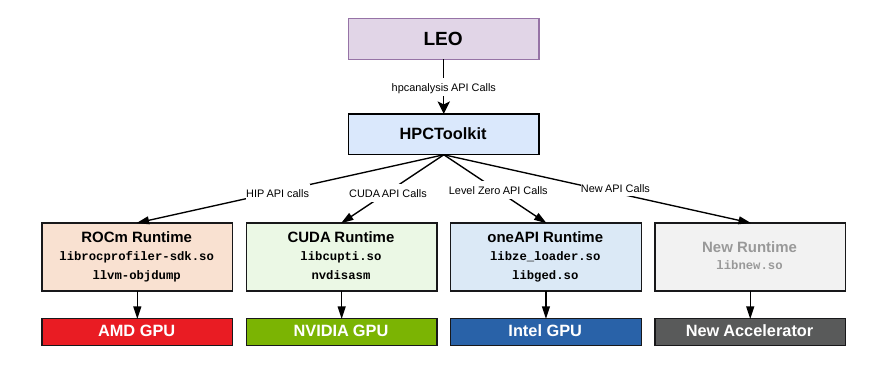}
\caption{System architecture. LEO interfaces with HPCToolkit's \tool{hpcanalysis} API, which leverages vendor-specific APIs for profiling and disassembly. The modular design enables extension to new accelerators.}
\label{fig:architecture}
\vspace{-1em}
\end{figure}

We evaluate LEO on 15 RAJAPerf~\cite{rajaperf} kernels, the QuickSilver~\cite{quicksilver} and Kripke~\cite{kripke} proxy applications, the llama.cpp inference engine~\cite{llamacpp}, and kernels from HipKittens~\cite{hipkittens}.
Our case studies show that LEO exposes actionable bottlenecks missed by vendor tools and that the same kernel can have different root causes of stalls on different GPU architectures.

The rest of the paper presents background (\cref{sec:background}), LEO's backward slicing (\cref{sec:frontend}), an evaluation of its utility (\cref{sec:eval}), case studies (\cref{sec:casestudies}), related work (\cref{sec:related}), and our conclusions (\cref{sec:conclusion}).

\section{Background}
\label{sec:background}

LEO analyzes fine-grained PC-sampling measurements collected using vendor-specific profiling APIs.
Because stall taxonomies, synchronization mechanisms, and concurrency semantics differ across GPUs from different vendors, their characteristics shape what LEO can infer.
This section compares AMD, NVIDIA, and Intel PC sampling along three axes: sampling mechanism, stall classification, and concurrency semantics.

\textbf{Terminology.}
AMD GPUs schedule \emph{waves} on compute units (CUs) while
NVIDIA GPUs schedule \emph{warps} on streaming multiprocessors (SMs).
Intel GPUs schedule hardware threads on Xe Vector Engines.
We use ``warp/wave'' when referring to vendor-specific execution groups and ``scheduler'' for the unit that selects the next instruction to issue.
Vendors use distinct terminology for on-chip scratchpad memory:  AMD's \emph{Local Data Store} (LDS), Intel's \emph{Shared Local Memory} (SLM), and NVIDIA's \emph{shared memory} (SMEM). On AMD GPUs, vector general-purpose registers are called VGPRs and high register usage reduces \emph{occupancy}---the number of concurrent waves/warps per CU/SM.

\subsection{PC Sampling on NVIDIA GPUs}
\label{sec:bg-nvidia}

NVIDIA's PC sampling periodically selects an active warp on each streaming multiprocessor (SM) and records its instruction address along with the warp scheduler state.
Each sample captures whether the warp issued an instruction or was stalled; stalls are classified into 13 categories defined by CUPTI (CUDA Profiling Tools Interface)~\cite{cupti2024}, including instruction fetch, execution dependency, memory dependency, texture, synchronization, constant memory dependency, pipe busy, memory throttle, not selected, and sleeping.
For devices with compute capability 6.0+ (P100 and later), NVIDIA additionally distinguishes between \emph{latency samples} (where a stall prevented issue) and total \emph{samples} (including cycles where the warp issued an instruction successfully), enabling more precise stall characterization.

NVIDIA exposes this capability through two CUPTI interfaces with different concurrency semantics.
The Activity API provides PC sampling on devices with compute capability 5.2+ (GTX 980 and later) but serializes kernel executions on the GPU during collection~\cite{cupti2024}, simplifying sample attribution but distorting measurements for applications that rely on concurrent kernels or streams. NVIDIA GPUs also support a lightweight PC sampling API that samples concurrent kernels without serialization but lacks support for correlating samples to kernel invocations.
For root cause tracing, NVIDIA's per-instruction stall categories provide the starting point: each stalled instruction's stall reason determines which type of dependency to trace backward.

\subsection{PC Sampling on AMD GPUs}
\label{sec:bg-amd}

AMD exposes PC sampling through the ROCprofiler-SDK API~\cite{rocprofiler2024}.
In \emph{stochastic} mode, available on MI300 and later, dedicated hardware periodically samples an active wave in each compute unit with no observable skid in our microbenchmarks.
Each sample reports the address and class of the sampled instruction, the instruction's stall reason (if any), an execution mask indicating active lanes in the sampled wave, and whether the wave issued. Stalls are classified as no instruction available, ALU dependency, waiting for memory, internal instruction, barrier wait, not selected, pipeline stall, sleep, and other.

Stochastic samples also report the count of active waves for the sampled cycle and per-pipeline arbiter bits, which indicate whether each pipeline issued and possibly stalled in the sampled cycle.
This pipeline-state visibility (unavailable on other vendor's GPUs) enables tools to distinguish an \emph{exposed} instruction stall (when the instruction's associated pipeline didn't issue in the sampled cycle) from a \emph{hidden} stall (when another instruction issued on the pipeline in the sampled cycle).
For root cause tracing, AMD's stochastic mode provides richer input than other vendors: stall reasons guide dependency tracing and pipeline state distinguishes whether identified stalls actually degrade performance or are hidden by wave parallelism.

\subsection{PC Sampling on Intel GPUs}
\label{sec:bg-intel}

Intel Data Center GPU Max (Ponte Vecchio) provides hardware-assisted periodic EU stall sampling exposed via Level Zero metrics interfaces~\cite{vtune2024,levelzero2024}. Each sample from an Xe Vector Engine (EU) consists of an instruction address and stall reason, if any.
Intel defines an execution unit as ``stalled'' when at least one thread is loaded but no thread can execute in the sampled cycle.
Stall categories include control flow stalls, pipeline hazards, memory send operations, scoreboard ID dependencies (SbidStall), synchronization, instruction fetch, distribution stalls, and other stalls~\cite{gtpin2024}. 

Unlike NVIDIA's Activity API, Intel's sampling preserves natural kernel concurrency.
Vendor documentation reports approximately \qty{10}{\percent} overhead when using the shortest recommended sampling period of \qty{100}{\micro\second}, making hardware sampling practical for production HPC workloads.
EU stall sampling is only supported on Intel's Data Center GPUs.
For root cause tracing, Intel's scoreboard ID (SbidStall) category is particularly informative: it indicates synchronization dependencies between send (memory) instructions and their consumers, directly guiding dependency edge construction.

\subsection{Summary and Implications}
\label{sec:bg-summary}

Three observations from this comparison guided our design.
First, stall taxonomies are vendor-specific: NVIDIA reports 13 categories, AMD stochastic mode reports 10+, and Intel reports 8.
While the categories overlap conceptually (memory, synchronization, dependencies), their definitions differ, requiring LEO to map vendor-specific stall reasons to a common dependency classification during root cause tracing.
Second, concurrency semantics vary significantly: NVIDIA's Activity API serializes kernels, while AMD and Intel preserve concurrency.
Third, AMD uniquely exposes hidden latency through pipeline state visibility, enabling distinction between exposed stalls and stalls hidden by wave parallelism.

\section{Backward Slicing through GPU Machine Code}
\label{sec:frontend}

While PC sampling reveals \emph{where} stalls occur, it does not explain \emph{why}.
LEO addresses this gap through backward slicing~\cite{cifuentes1997intraprocedural,srinivasan2016improved}: starting from a stalled instruction, LEO slices backward over instruction-level register, predicate, and vendor-specific synchronization dependencies, then applies pruning and blame attribution to identify the root cause instructions responsible for observed latency.
LEO builds on the machine-code backward-slicing approaches of Cifuentes and Fraboulet~\cite{cifuentes1997intraprocedural} and Srinivasan and Reps~\cite{srinivasan2016improved}, extending them with GPU-specific pruning heuristics and blame attribution for stall diagnosis.
This section describes LEO's analysis workflow, which also builds on the approach pioneered by GPA~\cite{zhou2021gpa} with cross-vendor support and enhanced synchronization tracing.

\Cref{lst:kernel} shows RAJAPerf's LTIMES\_NOVIEW kernel, a fused multiply-add involving three arrays.
\begin{figure}[tb]
\begin{lstlisting}[language=C++,
  xleftmargin=15pt, framexleftmargin=0pt,
  morekeywords={Index_type}]
if (m < num_m && g < num_g && z < num_z) {
 for (Index_type d = 0; d < num_d; ++d) {
  phidat[m+g*num_m+z*num_m*num_g] +=
   elldat[d+m*num_d] *
   psidat[d+g*num_d+z*num_d*num_g];
 }
}
\end{lstlisting}
\caption{RAJAPerf's LTIMES\_NOVIEW kernel: fused multiply-add involving three arrays.}
\label{lst:kernel}
\vspace{-1em}
\end{figure}

\begin{figure*}[t]
\centering
\includegraphics[width=\textwidth]{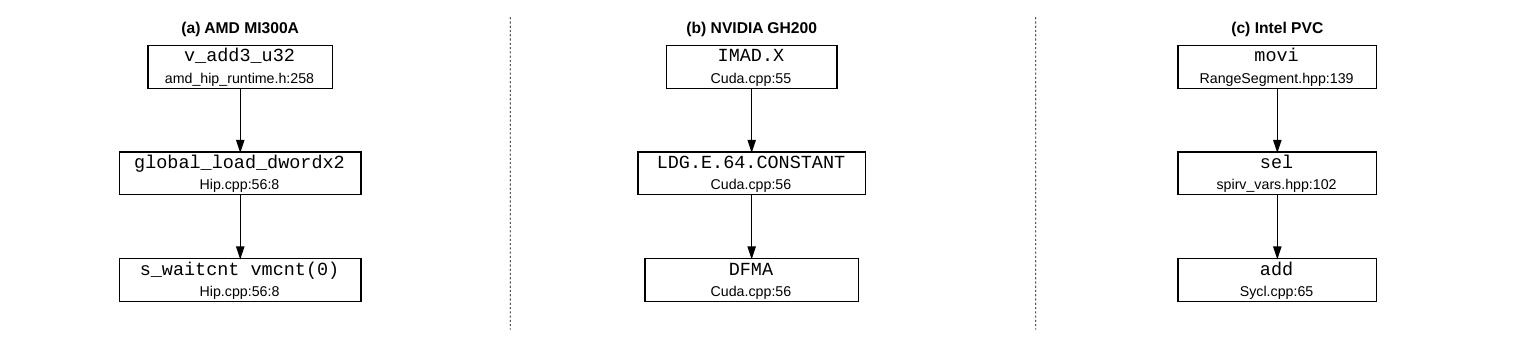}
\caption{Backward slicing on RAJAPerf's LTIMES\_NOVIEW kernel (\cref{lst:kernel}) across AMD, NVIDIA, and Intel GPUs. The same strided memory access bottleneck produces different dependency chains on each architecture, with root cause instructions tracing to different source files on AMD and Intel GPUs.}
\label{fig:backslicing-workflow}
\vspace{-1em}
\end{figure*}

\noindent \Cref{fig:backslicing-workflow} illustrates the LTIMES\_NOVIEW kernel on AMD, NVIDIA, and Intel GPUs.
Although the dependency chains differ across platforms, LEO traces them back to the same underlying cause: strided global-memory accesses to \code{elldat}.
This example shows why cross-vendor root-cause analysis matters: some kernels require architecture-specific fixes, whereas others expose a shared bottleneck that can only be confirmed through unified analysis.

\subsection{Workflow Overview}
\label{sec:backslicing-workflow}

LEO's analysis follows a 5-phase workflow:

\begin{enumerate}[nosep,leftmargin=*]
  \item \textbf{Data collection.} Disassembling the GPU binary (\tool{nvdisasm} for NVIDIA, \tool{llvm-objdump} for AMD, GED for Intel) and reading the HPCToolkit profile database via the \tool{hpcanalysis} API~\cite{grbic2025analyzing}.
  \item \textbf{Binary analysis.} Parsing binary sections, extracting instructions with register operands, and building control-flow graphs~\cite{meng2021parallel} with DWARF debug info for source mapping.
  \item \textbf{Dependency-graph construction.} Building a calling-context-tree (CCT) dependency graph from register dataflow (\cref{sec:backslicing-graph}) and extending it with vendor-specific synchronization tracing (\cref{sec:backslicing-sync}).
  \item \textbf{Four-stage pruning.} Filtering dependencies through opcode, barrier, latency, and execution constraints (\cref{sec:backslicing-pruning}).
  \item \textbf{Blame attribution.} Distributing stall cycles to surviving root causes using inverse-distance weighting (\cref{sec:backslicing-blame}).
\end{enumerate}
Because blame attribution relies on heuristic weighting rather than formal verification, LEO's root-cause reports should be understood as actionable attributions supported by dependence analysis and sampled stall evidence.

\subsection{CCT Dependency Graph}
\label{sec:backslicing-graph}

HPCToolkit attributes per-instruction performance metrics such as stall cycles by category to instruction offsets within each kernel. HPCToolkit organizes information about a kernel's instructions into a Calling Context Tree (CCT) that may include calls to device functions, inlined functions and templates, loops and statements.
LEO constructs a dependency graph between instructions annotated with stall cycle counts that are associated with nodes in this CCT.
Edges represent register RAW (read-after-write) hazards and point \emph{backward} in execution: from a stalled instruction (effect) to the instruction(s) that may have produced its source operand(s) (cause).

LEO builds a control-flow graph (CFG) for each device function and computes reaching definitions for machine-register writes using standard forward dataflow fixed-point iteration over GEN/KILL sets. The analysis operates directly on disassembled machine code rather than on an SSA IR; at control-flow joins, it unions reaching-definition sets from each incoming edge.

To obtain per-use precision, LEO performs a second, instruction-by-instruction forward walk through each basic block. Reaching definitions for the first instruction in a block are the block's incoming set. Each definition of a register within a block kills incoming definitions or prior definitions in the block. LEO links each use of a source operand to its set of reaching definitions. LEO then uses a backward liveness pass as a conservative cross-block filter: if a register is not live out of a defining block, LEO removes that candidate dependency. LEO then creates backward edges for each register used by an instruction to other instructions that may produce a reaching register value. When a producer instruction has no profile samples, LEO still retains it as an unsampled dependency source during graph construction so address-generation or predicate-setting instructions can receive blame.
The graph tracks dependencies across general-purpose registers (vector and scalar), predicate registers, barrier registers (NVIDIA B1--B6), and uniform registers (AMD scalar).
For predicated instructions, LEO tracks guard predicates (P0--P6 on NVIDIA) alongside data registers, ensuring that conditional definitions are included in the dependency graph even when only a subset of threads executes the defining instruction.

\begin{figure}[t]
\centering
\resizebox{\columnwidth}{!}{%
$\displaystyle
\begin{array}{@{}c@{\qquad}c@{}}
\begin{array}{l}
{\color{codekw}{\mathbf{for}}}\; \text{each basic block } B\text{:} \\
\quad \mathit{curr} \leftarrow \mathit{reach\_in}[B] \\
\quad {\color{codekw}{\mathbf{for}}}\; \text{each instruction } i \in B\text{:} \\
\quad\quad {\color{codekw}{\mathbf{for}}}\; \text{each source register } r\text{:} \\
\quad\quad\quad {\color{codekw}{\mathbf{for}}}\; \text{each } d \in \mathit{curr}[r]\text{:} \\
\quad\quad\quad\quad \text{add\_edge}(d \to i,\ \text{RAW}) \\[4pt]
\quad\quad {\color{codekw}{\mathbf{if}}}\; \text{predicated}(i)\text{:} \\
\quad\quad\quad {\color{codekw}{\mathbf{for}}}\; \text{each } d \in \mathit{curr}[\mathit{pred}(i)]\text{:} \\
\quad\quad\quad\quad \text{add\_edge}(d \to i,\ \text{GUARD}) \\[4pt]
\quad\quad {\color{codekw}{\mathbf{for}}}\; \text{each destination } w\text{:}\;
\mathit{curr}[w] \leftarrow \{i\}
\end{array}
&
\vcenter{\hbox{%
\begin{tikzpicture}[>=Stealth, node distance=0.9cm, every node/.style={font=\scriptsize}]
  \node[draw,rounded corners,fill=blue!8] (ld) {\texttt{LDG.E.64}};
  \node[draw,rounded corners,fill=blue!8,below left=0.8cm and 0.55cm of ld] (mul) {\texttt{DMUL}};
  \node[draw,rounded corners,fill=blue!8,below right=0.8cm and 0.55cm of ld] (setp) {\texttt{ISETP}};
  \node[draw,rounded corners,fill=red!10,below=0.8cm of mul] (fma) {\texttt{@P0 DFMA}};

  \draw[->,thick] (ld) -- (mul) node[midway,left,font=\tiny] {\texttt{R0}};
  \draw[->,thick] (ld) -- (setp) node[midway,right,font=\tiny] {\texttt{R1}};
  \draw[->,thick] (mul) -- (fma) node[midway,left,font=\tiny] {\texttt{R2}};
  \draw[->,thick,dashed] (setp) -- (fma) node[midway,right,font=\tiny] {\texttt{P0}};
\end{tikzpicture}%
}}
\end{array}
$
}
\caption{Dependency-graph construction (illustrated with NVIDIA SASS; the same algorithm applies to AMD and Intel). Left: intra-block edge construction from the block-entry reaching-definition set $\mathit{reach\_in}[B]$. Right: a single-block example. Solid arrows are register RAW dependencies; the dashed arrow is a predicate guard dependency.}
\label{fig:dep-graph}
\vspace{-1em}
\end{figure}

\subsection{4-Stage Pruning Pipeline}
\label{sec:backslicing-pruning}

The initial dependency graph is conservative and therefore contains many irrelevant edges.
LEO applies four sequential pruning stages to remove false dependencies:

\begin{enumerate}[nosep,leftmargin=*]
  \item \textbf{Opcode Constraints.}
  Edges are pruned based on compatibility between the source instruction's type and the destination's stall profile.
  If the destination shows only memory stalls, edges from compute instructions are removed; if it shows only execution dependency stalls, edges from global memory loads are removed.
  Edges constructed by synchronization tracing (\cref{sec:backslicing-sync}) are exempt from this stage, as they represent compiler-verified dependencies.

  \item \textbf{Barrier Constraints.}
  NVIDIA GPUs use numbered hardware barriers (B1--B6) encoded in each instruction's control field.
  An edge is removed if the source instruction sets a barrier that the destination does not wait on.
  This stage applies only to NVIDIA; AMD and Intel use different synchronization mechanisms described in \cref{sec:backslicing-sync}.

  \item \textbf{Latency Constraints.}
  If enough issue cycles separate a producer from its consumer, the dependency latency is hidden by the pipeline.
  LEO traverses CFG paths from producer to consumer, accumulating \metric{control.stall} cycles (NVIDIA) or instruction counts (AMD/Intel) at each instruction.
  An edge is pruned if the accumulated issue cycles exceed the producer's latency threshold on \emph{all} CFG paths.
  Valid (non-hidden) paths are stored on each edge for distance computation during blame attribution.

  \item \textbf{Execution Constraints.}
  Edges from instructions with zero execution count are optionally pruned.
\end{enumerate}

\subsection{Blame Attribution}
\label{sec:backslicing-blame}

After pruning, LEO distributes stall cycles from each stalled instruction to its surviving dependencies using four-factor weighting:

\begin{equation}
\text{blame}_i = S_j \times \frac{\mathcal{R}_i^{\text{dist}} \times \mathcal{R}_i^{\text{eff}} \times \mathcal{R}_i^{\text{isu}} \times \mathcal{R}_i^{\text{match}}}{\sum_{k} \mathcal{R}_k^{\text{dist}} \times \mathcal{R}_k^{\text{eff}} \times \mathcal{R}_k^{\text{isu}} \times \mathcal{R}_k^{\text{match}}}
\label{eq:blame-distribution}
\end{equation}

where $S_j$ is the total stall cycles at the stalled instruction $j$, and the four factors for each incoming dependency $i$ are:

\begin{itemize}[nosep,leftmargin=*]
  \item $\mathcal{R}_i^{\text{dist}} = d_{\min} / d_i$: \textbf{Distance factor.} Closer instructions receive more blame, as they have less opportunity to hide latency. $d_i$ is the average instruction count across valid CFG paths from Stage~3.
  \item $\mathcal{R}_i^{\text{eff}} = e_{\min} / e_i$: \textbf{Efficiency factor.} Less efficient instructions (e.g., uncoalesced memory accesses) receive more blame.
  \item $\mathcal{R}_i^{\text{isu}} = n_i / \sum_k n_k$: \textbf{Issue factor.} Instructions executed more frequently receive proportionally more blame.
  \item $\mathcal{R}_i^{\text{match}}$: \textbf{Stall-category match factor.} Weights each edge by how well its dependency type (memory, execution, synchronization) matches the destination's hardware-reported stall breakdown. For example, a memory dependency edge receives weight 0.8 if \qty{80}{\percent} of the destination's stall cycles are memory stalls.
\end{itemize}

The first three factors follow GPA's design~\cite{zhou2021gpa}; the fourth is a LEO extension that grounds blame in observed hardware metrics rather than purely static analysis.

When no dependencies survive pruning, LEO classifies the instruction as \emph{self-blame} with a diagnostic subcategory (memory latency, compute saturation, synchronization overhead, pipeline contention, instruction fetch, or indirect addressing) based on the dominant stall type in the hardware profile.

The proximate cause of a stall is the instruction that is stalled; a root cause is an earlier instruction that caused the stall.
In practice, LEO's chains often traverse framework layers (e.g., RAJA iterators, SYCL accessors), guiding developers from the stalled instruction to the actionable design decision.
This definition is consistent with instruction-level diagnosis tools such as GPA~\cite{zhou2021gpa}.

\subsection{Cross-Vendor Synchronization Tracing}
\label{sec:backslicing-sync}

Purely register-based tracing breaks at synchronization instructions because they expose no explicit register dependencies.
LEO avoids these dead ends by adding vendor-specific synchronization edges that bypass opcode and latency pruning:

\textbf{AMD: \instr{s\_waitcnt} tracing.}
AMD's \instr{s\_waitcnt} instruction waits until counts of in-flight memory operations drain to a specified level. Counts include \metric{vmcnt} for vector memory operations (\instr{global\_load}, \instr{buffer\_load}) and \metric{lgkmcnt} for LDS/constant operations (\instr{ds\_read}, \instr{s\_load}).
When LEO encounters \instr{s\_waitcnt vmcnt(N)}, it scans backward to find the $(M{-}N)$ oldest pending memory operations (where $M$ is the total count), stopping at epoch boundaries where a prior \instr{s\_waitcnt} already drained the counter.
These operations are added as \metric{mem\_waitcnt} edges.

\textbf{NVIDIA: barrier tracing.}
NVIDIA encodes hardware barriers (B1--B6) in each instruction's control field via \metric{Control.read}/\metric{Control.write} bits.
When an instruction waits on a barrier (via \metric{Control.wait} bitmask or \instr{DEPBAR} operand), LEO scans backward to find instructions that set matching barriers, creating \metric{mem\_barrier} edges.

\textbf{Intel: SWSB token tracing.}
Intel Xe HPC uses SWSB tokens (SBID 0--31) for dependency synchronization.
When an instruction contains a wait directive (\instr{dst\_wait} or \instr{src\_wait} on SBID $T$), LEO scans backward to find the \instr{send} instruction that set SBID $T$, creating \metric{mem\_swsb} edges.

\textbf{Unified framework.}
LEO provides a unified interface for tracing memory dependencies. It accommodates differences in GPU synchronization mechanisms that await completion of memory accesses by dispatching to a vendor-specific algorithm that produces typed edges (\metric{mem\_waitcnt}, \metric{mem\_barrier}, \metric{mem\_swsb}) that are exempt from opcode and latency pruning.
This extended tracing exceeds GPA's capability: GPA traces NVIDIA barriers but lacks support for AMD and Intel GPUs, which employ different mechanisms for awaiting completion of memory accesses.

\section{Large Language Model-Based Optimization}
\label{sec:llm}

Beyond aiding manual optimization, we evaluate whether results from LEO's root-cause analysis improve large language model (LLM)-based code optimization.
Our workflow for LLM-based kernel tuning has five stages: (1)~HPCToolkit collects PC samples; (2)~LEO performs root-cause analysis; (3)~LEO formats the result as a structured stall report with dependency chains and source mappings; (4)~a two-stage LLM pipeline uses a \emph{strategist} to propose optimization strategies and a \emph{code generator} to implement them; and (5)~the generated code is compiled, verified, and benchmarked.

\subsection{Diagnostic Context for LLMs}
\label{sec:llm-context}

Profiles alone tell an LLM \emph{where} time is spent, but not \emph{why}, so it mainly supports generic optimization heuristics.
LEO adds three forms of diagnostic context: \emph{root-cause identification} (e.g., ``line 56 stalls because of strided \texttt{elldat} accesses'' rather than ``line 56 is slow''), \emph{cross-file dependency chains} that expose the critical path, and \emph{quantified impact} via cycle counts that help prioritize high-value changes.

\subsection{LLM Diagnostic Context Comparison}
\label{sec:llm-context-study}

To isolate LEO's contribution, we compare three diagnostic settings: \textbf{C} (code only), \textbf{C+S} (code plus raw per-instruction stall counts), and \textbf{C+L(S)} (code plus LEO's full root-cause analysis, including dependency chains and blame attribution).
Comparing C with C+S tests whether stall profiles help. Comparing C+S with C+L(S) isolates the value of causal analysis.
This study complements recent work on LLM-based parallel code generation~\cite{nichols2024can}, performance-improving code edits~\cite{shypula2024learning}, and performance-tool-guided kernel optimization~\cite{nichols2025perftoolsllm,tschand2025swizzleperf} by evaluating whether structured profiler diagnostics can guide such models more effectively.
We evaluate Gemini 3.1 Pro on 15 RAJAPerf CUDA kernels with $k{=}5$ trials and temperature 0.7 (\cref{sec:eval-ablation}).
Multi-model evaluation and expanded kernel coverage remain future work.

\section{Evaluation}
\label{sec:eval}

This section evaluates LEO across three GPU platforms, focusing on the quality of root cause analysis, optimization impact, and utility for automated tuning.
Our evaluation focuses on whether LEO provides useful and actionable attributions for optimization, rather than on formally verifying the causal correctness of every reported chain.

\subsection{Experimental Setup}
\label{sec:eval-setup}

\paragraph{Hardware Platforms}
\Cref{tab:eval-hardware} summarizes the three AMD, NVIDIA, and Intel GPUs used in our evaluation.

\begin{table}[!t]
\centering
\caption{GPU hardware platforms used for evaluation.}
\label{tab:eval-hardware}
\small
\begin{adjustbox}{max width=\columnwidth}
\begin{tabular}{@{}llll@{}}
\toprule
& \textbf{AMD MI300A} & \textbf{NVIDIA GH200} & \textbf{Intel GPU Max 1100} \\
\midrule
Architecture & CDNA3 APU & Hopper (Grace Hopper) & Xe-HPC (Ponte Vecchio) \\
Compute Units & \num{228}~CUs & \num{132}~SMs & \num{56}~Xe-cores (\num{448}~XVEs) \\
Memory & \qty{128}{\giga\byte} HBM3 & \qty{480}{\giga\byte} HBM3e & \qty{48}{\giga\byte} HBM2e \\
FP64 Peak & \qty{61.3}{\TFLOPS} & \qty{33.5}{\TFLOPS} & \qty{22.2}{\TFLOPS} \\
Mem BW & \qty{5.3}{\tera\byte\per\second} & \qty{4.0}{\tera\byte\per\second} & \qty{1.23}{\tera\byte\per\second} \\
Software & ROCm 6.3 & CUDA 12.8 & Level Zero 1.x \\
\bottomrule
\end{tabular}
\end{adjustbox}
\end{table}

\paragraph{HPC Workloads}
\Cref{tab:eval-workloads} summarizes the workloads.
RAJAPerf contributes 15 kernels extracted from production LLNL codes; for the cross-platform root-cause study, we analyze and optimize the \code{Base\_CUDA}, \code{Base\_HIP}, and \code{Base\_SYCL} variants, and report speedup relative to the original version of each variant (\cref{tab:eval-backslice}).
The remaining applications support the case studies in \cref{sec:casestudies}.

\begin{table}[!t]
\centering
\caption{HPC and ML workloads used for evaluation.}
\label{tab:eval-workloads}
\small
\begin{adjustbox}{max width=\columnwidth}
\begin{tabular}{@{}lllll@{}}
\toprule
\textbf{Workload} & \textbf{Domain} & \textbf{Type} & \textbf{Prog.\ Model} & \textbf{Developer} \\
\midrule
RAJAPerf~\cite{rajaperf} & Multi-physics & Benchmark suite & CUDA/HIP/SYCL & LLNL \\
XSBench~\cite{tramm2014xsbench} & Nuclear physics & Proxy app & OpenMP offload & ANL \\
miniBUDE~\cite{poenaru2021minibude} & Drug discovery & Mini-app & CUDA/HIP/SYCL & U.\ Bristol \\
LULESH~\cite{karlin2013lulesh} & Hydrodynamics & Proxy app & Kokkos & LLNL \\
QuickSilver~\cite{quicksilver} & Particle transport & Proxy app & CUDA/HIP & LLNL \\
HipKittens~\cite{hipkittens} & ML kernels & Library & HIP & Stanford \\
llama.cpp~\cite{llamacpp} & LLM inference & Production & CUDA/HIP & Community \\
Kripke~\cite{kripke} & Particle transport & Proxy app & RAJA (CUDA/HIP) & LLNL \\
\bottomrule
\end{tabular}
\end{adjustbox}
\end{table}

\begin{table*}[!t]
\caption{Root cause analysis and optimization results across three GPU platforms. Per-kernel speedups measured by vendor GPU profilers (Nsight Systems, rocprofv3, unitrace).}
\label{tab:eval-backslice}
\centering
\setlength{\tabcolsep}{0.6pt}
\renewcommand{\arraystretch}{0.95}
\fontsize{5.4}{6.5}\selectfont
\begin{tabular}{@{}ll ccc c ccc c ccc@{}}
\toprule
& & \multicolumn{3}{c}{\textbf{NVIDIA GH200}} & & \multicolumn{3}{c}{\textbf{AMD MI300A}} & & \multicolumn{3}{c}{\textbf{Intel PVC}} \\
\cmidrule{3-5} \cmidrule{7-9} \cmidrule{11-13}
\textbf{Application} & \textbf{Kernel} & \textbf{Root Cause} & \textbf{Optimization} & \textbf{Speedup} & & \textbf{Root Cause} & \textbf{Optimization} & \textbf{Speedup} & & \textbf{Root Cause} & \textbf{Optimization} & \textbf{Speedup} \\
\midrule
\multicolumn{13}{@{}l}{\textit{RAJAPerf Kernels (baseline: \texttt{Base\_\{CUDA,HIP,SYCL\}})}} \\[2pt]
MASS3DEA           & \texttt{Mass3DEA}            & FP64 FMA Chain     & Precompute basis in regs        & 3.66$\times$  & & Register Spilling  & Precompute basis in regs        & 2.51$\times$ & & SLM Fence Latency  & Precompute basis in regs        & 10.32$\times$ \\
LTIMES\_NOVIEW     & \texttt{ltimes\_noview}      & Stride-64 Loads    & Tile \texttt{elldat} into SMEM   & 4.41$\times$  & & Uncoalesced Loads  & Tile \texttt{elldat} into LDS          & 4.89$\times$ & & Predication Overhead & Tile \texttt{elldat} into SLM          & 3.06$\times$ \\
LTIMES             & \texttt{ltimes}              & Stride-64 Loads    & Tile \texttt{elldat} into SMEM   & 5.02$\times$  & & Global Load Latency & Tile \texttt{elldat} into LDS          & 4.00$\times$ & & URB Contention     & Tile \texttt{elldat} into SLM          & 3.01$\times$ \\
3MM                & \texttt{poly\_3mm\{1--3\}}           & Global Load Latency & Tile A,B into SMEM   & 1.91$\times$  & & Global Load Latency & Tile A,B into LDS          & 3.38$\times$ & & URB Contention     & Tile A,B into SLM          & 2.98$\times$ \\
2MM                & \texttt{poly\_2mm\{1,2\}}           & Global Load Latency & Tile A,B into SMEM   & 1.69$\times$  & & Global Load Latency & Tile A,B into LDS          & 3.37$\times$ & & URB Contention     & Tile A,B into SLM          & 2.98$\times$ \\
GEMM               & \texttt{poly\_gemm}          & Global Load Latency & Tile A,B into SMEM   & 1.55$\times$  & & Global Load Latency & Tile A,B into LDS          & 3.26$\times$ & & URB Contention     & Tile A,B into SLM          & 2.98$\times$ \\
PRESSURE           & \texttt{pressurecalc\{1,2\}}     & Inter-Kernel Traffic & Fuse 2 kernels (reg bvc)       & 2.55$\times$  & & Inter-Kernel Traffic & Fuse 2 kernels (reg bvc)       & 2.06$\times$ & & Transcendental Stall & Fuse 2 kernels (reg bvc)       & 1.84$\times$ \\
ENERGY             & \texttt{energycalc\{1--6\}}  & Inter-Kernel Traffic & Fuse 6 kernels into 1       & 1.81$\times$  & & Inter-Kernel Traffic & Fuse 6 kernels into 1       & 2.34$\times$ & & Control Flow Stall & Fuse 6 kernels into 1       & 3.66$\times$ \\
FIR                & \texttt{fir}                 & FP64 FMA Chain     & Tile coefficients into SMEM        & 1.00$\times$           & & Global Load Stall  & Tile coefficients into LDS          & 1.86$\times$ & & URB Contention     & Tile coefficients into SLM          & 1.00$\times$ \\
ZONAL\_ACCUM\_3D   & \texttt{zonal\_accum\_3d} & Constant Load Stall & \texttt{\_\_ldg} + \texttt{\_\_restrict\_\_}        & 1.07$\times$   & & Scalar Load Stall  & 8 ptrs $\to$ base+arith   & 1.16$\times$ & & Memory Dep Stall   & 8 ptrs $\to$ base+stride   & 1.17$\times$ \\
VOL3D              & \texttt{vol3d}               & Pointer Indirection & 24 ptrs $\to$ 3 base+stride   & 1.06$\times$   & & VGPR Spill         & 24 ptrs $\to$ 3 base+stride   & 1.35$\times$ & & Control Flow Stall & 24 ptrs $\to$ 3 base+stride   & 1.17$\times$ \\
DEL\_DOT\_VEC\_2D  & \texttt{deldotvec2d}         & Reduction Load Stall & 16 ptrs $\to$ 4 base+stride   & 1.01$\times$   & & Reduction Stall    & 16 ptrs $\to$ 4 base+stride   & 1.03$\times$ & & URB Contention     & 16 ptrs $\to$ 4 base+stride   & 1.33$\times$ \\
DIFFUSION3DPA      & \texttt{Diffusion3DPA}       & Stencil Cache Miss & Tile basis B,G into SMEM   & 1.01$\times$  & & Stencil LDS Stall  & Tile basis B,G into LDS          & 1.07$\times$ & & URB Contention     & Tile basis B,G into SLM          & 1.00$\times$ \\
CONVECTION3DPA     & \texttt{Convection3DPA}      & Global Load Latency & Tile basis into SMEM   & 1.00$\times$  & & Stencil LDS Stall  & Tile basis into LDS          & 1.21$\times$ & & Predication Overhead & Tile basis into SLM          & 1.00$\times$ \\
MASS3DPA           & \texttt{Mass3DPA}            & Global-SLM Reload  & Cache B,Bt in SMEM         & 1.05$\times$   & & Low Occupancy (VGPR) & Occupancy hint + no unroll   & 1.02$\times$ & & SLM Fence Latency  & Merge phases (SLM)       & 1.08$\times$ \\
\midrule
\multicolumn{13}{@{}l}{\textit{Kernels from Applications or Mini-Apps}} \\[2pt]
LULESH             & \texttt{CalcFBHourglass}     & Scattered Node Access & Gather-once AoS$\to$SoA        & 1.00$\times$           & & Scattered Node Access & Gather-once AoS$\to$SoA  & 1.01$\times$ & & Scattered Node Access    & Gather-once AoS$\to$SoA        & 1.14$\times$ \\
miniBUDE           & \texttt{fasten\_main}        & Irregular Loads    & \texttt{\_\_ldg} float4 + restrict & 2.20$\times$ & & Irregular Loads    & restrict + unroll   & 1.20$\times$ & & Control Flow Stall & USM + work-group tune        & 1.03$\times$ \\
XSBench            & \texttt{xs\_lookup\_kernel}   & Irregular Loads    & Integer hash + extract   & 1.08$\times$ & & Irregular Loads    & Integer hash + extract   & 1.01$\times$ & & Irregular Loads    & Hybrid search + USM       & 1.18$\times$ \\
Kripke             & \texttt{LTimes}              & Global Load Stall  & Swap Zone$\leftrightarrow$Group (RAJA)        & 11.67$\times$ & & Global Load Stall  & Swap Zone$\leftrightarrow$Group (RAJA)        & 1.90$\times$ & & \multicolumn{3}{c}{N/A} \\
llama.cpp          & \texttt{mul\_mat\_q}         & Tensor Core Stall  & Direct store + restrict        & 1.00$\times$ & & VGPR Pressure (0\%) & Tile 128$\to$64 + direct store     & 1.12$\times$ & & \multicolumn{3}{c}{N/A} \\
QuickSilver        & \texttt{CycleTracking}       & Cross-File Ptr Chain & Inline cross-file call      & 1.51$\times$ & & Flat Load Latency  & Inline cross-file call      & 1.27$\times$ & & \multicolumn{3}{c}{N/A} \\
\midrule
\multicolumn{2}{@{}l}{\textbf{Geomean}} & & & \textbf{1.73$\times$} & & & & \textbf{1.74$\times$} & & & & \textbf{1.82$\times$} \\
\bottomrule
\end{tabular}
\end{table*}

\paragraph{Profiling and Analysis Overhead}
LEO performs post mortem analysis of previously collected profiles and therefore adds no runtime overhead.
Any runtime cost comes from HPCToolkit's hardware-assisted PC-sampling infrastructure. On AMD GPUs, PC sampling at HPCToolkit's default frequency adds  roughly 10\% measurement overhead to an execution.
LEO's analysis time is dominated by HPCToolkit's \tool{hpcanalysis} database reader.
After that, dependency-graph construction, pruning, and blame attribution typically finish in \qtyrange{3}{10}{\second} per kernel on one CPU core (e.g., \qty{3.6}{\second} for RAJAPerf on AMD and \qty{8.1}{\second} for QuickSilver on NVIDIA), although NVIDIA tensor-core kernels in llama.cpp with more than \num{8000} dependency edges and long scheduling chains took about \qty{60}{\second}.

\subsection{Dependency Tracing Effectiveness}
\label{sec:eval-backslice}

We evaluate LEO at two levels: diagnostic usefulness and optimization outcome.
For each RAJAPerf kernel, LEO produces a ranked stall chain through backward slicing. We then applied a minimal source-level modification confined to the code region implicated by the top-ranked chain (e.g., shared-memory tiling, kernel fusion, or pointer reduction) and re-measure performance using the same harness, problem size, and iteration count as the baseline kernel.
This protocol is intentionally restrictive: it does not allow unrelated algorithmic changes or broad manual retuning, so the resulting speedups provide a conservative measure of LEO's usefulness as a diagnostic tool.
RAJAPerf includes internal warmup phases and kernel-specific repetition counts.
Except for kernel-fusion cases (PRESSURE, ENERGY), we report speedup from per-kernel GPU execution time measured by vendor profilers: Nsight Systems on NVIDIA, rocprofv3 on AMD, and unitrace on Intel.
For PRESSURE and ENERGY, per-kernel GPU timings would miss the benefit of eliminating inter-kernel traffic and launch overhead, so we use RAJAPerf's built-in aggregate timer instead.
For case studies, we measured ten independent runs per configuration and report mean speedup.
For RAJAPerf kernels, we report the mean GPU kernel time across ten profiler invocations.
Run-to-run variability is generally low on NVIDIA and AMD (median CV below 1\%), with LULESH and Kripke showing higher variance. Intel PVC exhibits higher variability (median CV 9\%) due to platform-specific scheduling behavior.
We assessed statistical significance using two-sided paired t-tests on matched baseline/optimized timing samples. For the final measurements reported in \cref{tab:eval-backslice}, all speedups greater than 1.05$\times$ were significant at $p < 0.01$. Smaller changes near unity were mixed: some were statistically significant because run-to-run variance was very low, whereas others were not distinguishable from the baseline.
\Cref{tab:eval-backslice} presents the results across all three GPU platforms.

The results show two main findings.
First, LEO exposes \emph{cross-vendor root-cause divergence}: the same kernel can be compute-bound on one platform and memory-bound on another.
Across all 21 workloads, geometric-mean speedups are 1.73$\times$ on NVIDIA GH200, 1.74$\times$ on AMD MI300A, and 1.82$\times$ on Intel PVC.
Three codes (QuickSilver, llama.cpp, Kripke) lack SYCL ports; Intel results are omitted for these workloads.

\emph{Observation 1: The same kernel exhibits fundamentally different bottlenecks across GPU architectures.}
For example, FIR is compute-bound on NVIDIA (1.00$\times$) but memory-bound on AMD (1.86$\times$), and llama.cpp requires occupancy tuning on AMD (1.12$\times$) but has no dominant bottleneck on NVIDIA.

Second, LEO remains useful even when little speedup is available.
For DEL\_DOT\_VEC\_2D (near 1.0$\times$ on all platforms), LEO correctly attributes the bottleneck to reduction stalls with limited optimization headroom, steering developers away from unproductive effort.

\emph{Observation 2: Optimization transferability depends on bottleneck structure.}
Of 21 workloads, 11 benefit from the same optimization across vendors (e.g., shared-memory tiling for LTIMES, 3MM, GEMM; kernel fusion for PRESSURE, ENERGY), while six require vendor-specific fixes (e.g., miniBUDE, llama.cpp); the remaining four show mixed results or no actionable bottleneck.
Regular memory access patterns admit portable optimizations; architecture-specific stall mechanisms need customized solutions.

\subsection{Dependency Graph Quality}

To assess whether LEO produces graphs suitable for unambiguous blame attribution, we measure \emph{single-dependency coverage}~\cite{zhou2021gpa}: the fraction of nodes whose incoming edges belong to distinct dependency classes (e.g., memory versus execution), so blame can be assigned to one edge without apportionment.
\Cref{fig:sdc-coverage} shows coverage before and after LEO's analysis workflow (synchronization tracing + four-stage pruning) across all 21 workloads on three architectures.

\begin{figure*}[tb]
\centering
\resizebox{0.99\textwidth}{!}{\input{figures/sdc_coverage.pgf}}
\vspace{-1.5em}
\caption{Single-dependency coverage before (faded) and after (solid) LEO's analysis workflow. The dashed line marks \qty{80}{\percent}, above which blame is typically unambiguous. Orange bars indicate workloads where synchronization tracing increases graph complexity in exchange for deeper root-cause identification.}
\label{fig:sdc-coverage}
\vspace{-1em}
\end{figure*}

Without pruning, single dependency coverage is \qtyrange{30}{74}{\percent} on NVIDIA GH200. Pruning improves coverage  to \qtyrange{64}{94}{\percent}, with 13/21 workloads exceeding \qty{80}{\percent}.
On AMD MI300A, coverage reaches \qtyrange{80}{97}{\percent} for most workloads; three kernels (ENERGY, FIR, ZONAL\_ACCUM\_3D) show slight decreases because \texttt{s\_waitcnt} tracing adds memory counter edges that increase graph complexity---a necessary cost for tracing through synchronization barriers to reach actual memory operations.
On Intel PVC, coverage remains stable at \qtyrange{61}{89}{\percent} because Intel's SWSB mechanism produces compiler-verified dependency edges that are already precise, requiring minimal pruning.

Single dependency coverage measures the clarity of blame attribution---whether dependencies can be unambiguously classified---not its correctness.
We assess effectiveness through case studies where optimization guided by LEO's top-ranked chains achieves substantial speedup (\cref{sec:casestudies}).

\subsection{LLM Diagnostic Context Comparison}
\label{sec:eval-ablation}

As described in \cref{sec:llm-context-study}, we compare three levels of diagnostic context on NVIDIA GH200 using 15 RAJAPerf kernels and Gemini 3.1 Pro ($k{=}5$ trials, best-of-$k$ speedup).
\Cref{tab:ablation} presents the results.

\begin{table}[tb]
\caption{Diagnostic context comparison on NVIDIA GH200 using Gemini 3.1 Pro (15 RAJAPerf kernels, $k{=}5$ trials, best-of-$k$ speedup).
C: source code only. C+S: code plus raw stall counts. C+L(S): code plus LEO's root-cause analysis of stalls.}
\label{tab:ablation}

\footnotesize\centering
\setlength{\tabcolsep}{6pt}
\begin{tabular}{@{}lccc@{}}
\toprule
& \textbf{C} & \textbf{C+S} & \textbf{C+L(S)} \\
\midrule
Compilable (\%)       & \qty{37}{\percent}           & \qty{76}{\percent}                      & \textbf{\qty{100}{\percent}} \\
Geomean Speedup       & 1.13$\times$   & 1.08$\times$              & \textbf{1.29$\times$} \\
Regressions ($<$1$\times$) & 1         & 5 & \textbf{0} \\
\bottomrule
\end{tabular}
\vspace{-1em}
\end{table}

C+L(S) achieves a 1.29$\times$ geometric-mean speedup with all 15 kernels producing compilable code, compared with 1.13$\times$ (37\% compilable) for code-only input (C) and 1.08$\times$ (76\% compilable) for raw stalls (C+S).
Relative to C, LEO raises the fraction of compilable outputs from 37\% to 100\%. Five kernels that produce no compilable code under C achieve 1.26--1.97$\times$ speedup with LEO's guidance.
Notably, raw stall data alone can hurt optimization quality: PRESSURE drops to 0.85$\times$ under C+S (versus 1.20$\times$ with LEO), and VOL3D regresses to 0.36$\times$. These results suggest that unstructured profiling data can mislead LLMs, whereas causal dependency chains support targeted changes.

\emph{Observation 3: Structured dependency chains guide optimization better than raw metrics.}
Causal chains suppress false positives by linking stalls to actionable code regions: without them, LLMs apply generic transformations that can degrade performance (PRESSURE achieves 0.85$\times$ speedup and VOL3D achieves 0.36$\times$ speedup after transformations guided by raw stalls).

\section{Case Studies}
\label{sec:casestudies}

We present case studies that highlight LEO's diagnostic range, including cross-vendor root-cause divergence, architecture-specific optimizations, cross-file tracing, machine-learning kernels, and framework abstraction layers.

\subsection{MASS3DEA: One Optimization, Three Root Causes}
\label{sec:case-mass3dea}

RAJAPerf's MASS3DEA kernel computes 3D finite-element mass-matrix assembly through basis-function products.
LEO shows that the same kernel has very different stall reasons on the three GPU platforms. 
\begin{itemize}
    \item On NVIDIA GH200, FP64 multiply dependency chains (DMUL$\to$DMUL, \qty{52.3}{\percent} of stall cycles) limit instruction-level parallelism.
    \item On AMD MI300A, severe register pressure (\num{189}~VGPRs, \qty{0}{\percent} occupancy) shifts \qty{84}{\percent} of cycles to \instr{s\_waitcnt} synchronization.
    \item On Intel PVC, SLM-fence latency through SYCL accessor layers (\file{accessor.hpp}) accounts for \qty{28}{\percent} of stalls.
\end{itemize}  
Despite these different causes, the same optimization---precomputing basis-function values in registers---addresses all three cases. It breaks the dependency chain on NVIDIA, reduces register pressure on AMD, and removes SLM round trips on Intel.
The resulting speedups are 3.66$\times$, 2.51$\times$, and 10.32$\times$, respectively, with Intel benefiting most because the SLM-fence overhead dominates.

\subsection{miniBUDE: Architecture-Specific Optimizations}
\label{sec:case-minibude}

miniBUDE~\cite{poenaru2021minibude} is a molecular-docking mini-app with irregular memory accesses.
Although LEO identifies irregular loads as the common symptom on all three architectures, the underlying mechanisms differ, so the effective code changes are vendor-specific.
\begin{itemize}
    \item On NVIDIA GH200, \qty{20.8}{\percent} of stalls arise at \instr{LDG.E.U8.CONSTANT}, where address computation delays load issue. We therefore use \code{\_\_ldg()}, \code{float4} coalescing, and \code{\_\_restrict\_\_} to route accesses through the read-only cache, yielding 2.20$\times$.
    \item On AMD MI300A, \qty{24.5}{\percent} of stalls arise at \instr{s\_waitcnt vmcnt}; because all 64 lanes redundantly load data for the same protein atoms, we replace vector loads with scalar broadcasts (\instr{s\_load\_dwordx4}), yielding 1.20$\times$.
    \item On Intel PVC, \qty{52.4}{\percent} of sampled stalls come from control-flow overhead introduced by SYCL accessor indirection, so we switch to USM pointers and retune work-group sizes, yielding 1.03$\times$.
\end{itemize}
These optimizations are architecture-specific: no optimization applies directly across GPUs from different vendors.

\subsection{QuickSilver: Cross-File Root Cause Tracing}
\label{sec:case-quicksilver}

QuickSilver~\cite{quicksilver} is a Monte Carlo particle-transport proxy for LLNL's Mercury code.
\begin{itemize}
    \item On NVIDIA GH200, LEO traces a three-file dependency chain responsible for \qty{19.8}{\percent} of stall cycles: a floating-point multiply in \file{CollisionEvent.hh} stalls on a global load that passes through \file{MacroscopicCrossSection.hh} to \file{NuclearData.hh}.
    \item On AMD MI300A, LEO instead points to atomic operations and \instr{s\_waitcnt} fences under extreme register pressure (\num{153}~VGPRs, \qty{0}{\percent} occupancy).
\end{itemize}
Guided by these platform-specific diagnoses, we inline the cross-file call, add \code{\_\_restrict\_\_}, and hoist loop invariants on NVIDIA; on AMD, we reduce register pressure along the lookup path.
The resulting speedups are 1.51$\times$ on GH200 and 1.27$\times$ on MI300A.
The key point is that the root cause in \file{NuclearData.hh} is not visible from \file{CollisionEvent.hh} alone; only LEO's cross-file chain exposes it.

\subsection{ML Kernel Optimization: llama.cpp and HipKittens}
\label{sec:case-ml}

\paragraph{llama.cpp: Vendor-Specific Optimization}
llama.cpp~\cite{llamacpp} is a widely used large language model inference engine.
LEO identifies fundamentally different root causes in the same quantized matrix-multiply kernel (\file{mmq.cuh}) on AMD MI300A and NVIDIA GH200.

\begin{figure}[tb]
\begin{lstlisting}[language=C++,
  xleftmargin=15pt, framexleftmargin=0pt,
  keywordstyle=\ttm]
(*@\color{codecomment}{\textit{// mul\_mat\_q kernel: store results}}@*)
for (int j = j0; j < ncols; j++) {
    if (ids_dst[j] < 0) continue;
(*@\rminus@*) (*@\textcolor{red}{  dst[ids\_dst[j]*stride + i] = sum[...];}@*)
(*@\gplus@*) (*@\textcolor{darkgreen}{  dst[j*stride + i] = sum[...]; // direct}@*)
}
\end{lstlisting}
\caption{llama.cpp \texttt{mmq.cuh}: LEO identifies \texttt{ids\_dst[j]} as a serializing LDS read. Replacing the indirect store with direct indexing yields 1.12$\times$ on AMD MI300A.}
\label{lst:llama-diff}
\vspace{-0.5em}
\end{figure}

\begin{itemize}
\item
On AMD, LEO reports a \qty{70.9}{\percent} stall ratio and \qty{0}{\percent} occupancy, with \numrange{160}{256}~VGPRs per wave. 
\item
On NVIDIA, LEO finds no dominant stall hotspot; stall cycles are distributed across multiple pipeline stages with no single actionable bottleneck.
\end{itemize}
Guided by the AMD-specific diagnosis, we reduce tile width from 128 to 64 to restore occupancy and replace indirect stores with direct addressing (\cref{lst:llama-diff}), yielding 1.12$\times$ on AMD (Qwen2.5-1.5B, Q4\_K\_M quantization). This case illustrates cross-vendor divergence: the same kernel requires optimization on one platform but is already efficient on another.

\paragraph{HipKittens: Expert-Optimized ML Kernels}
HipKittens~\cite{hipkittens} shows that LEO can help even on hand-tuned kernels. Its RMSNorm kernel for AMD MI300A was already optimized by expert developers, yet LEO traces a cross-file chain showing that the compiler lowers BF16 vector loads to scalar \instr{global\_load\_ushort} operations, leaving \qtyrange{20}{58}{\percent} of stall cycles attributable to memory. Guided by that diagnosis, we implement multi-row software pipelining with split \instr{s\_waitcnt} counters and obtain 1.07$\times$--1.24$\times$ speedup on MI300A. This case shows that LEO can uncover additional opportunities even in code written by experts.

\subsection{Kripke: Tracing Through RAJA Abstraction Layers}
\label{sec:case-kripke}

Kripke~\cite{kripke} is an LLNL proxy application for deterministic particle transport implemented using the RAJA template-based programming model.
The physics kernel is a single line: \code{phi(nm,g,z) += ell(nm,d) * psi(d,g,z)}.
Nothing in the source code suggests a performance problem, yet LEO reports that this line accounts for \qty{96.7}{\percent} of stall cycles on NVIDIA GH200 (\cref{fig:kripke-chain}).

LEO's backward slice reveals why: the stalling \instr{DFMA} instruction waits on a global load whose address computation chains through three RAJA framework layers: array indexing in \file{TypedViewBase.hpp}, offset arithmetic in \file{Operators.hpp}, and strided iteration logic in \file{Iterators.hpp}.
The appearance of RAJA's strided iterator at the end of the chain is the key diagnostic clue: it tells the developer that the memory access pattern is determined by the RAJA execution policy, not the physics code.
Inspecting the RAJA execution policy in \file{LTimes.h} explains the strided pattern: mapping \code{Group} to \code{cuda\_thread\_x} while keeping \code{Zone} as a sequential inner loop forces adjacent GPU threads to access memory \qty{32768}{\byte} apart.

The fix is a two-line change in \file{LTimes.h}: swapping the two dimensions aligns adjacent threads with contiguous memory, yielding 11.67$\times$ on NVIDIA GH200 but only 1.90$\times$ on AMD MI300A (\cref{tab:eval-backslice}), reflecting MI300A's superior HBM3 bandwidth, which reduces the penalty for not coalescing.
On AMD, LEO traces through \instr{s\_waitcnt vmcnt(0)} to the same strided \instr{global\_load}, confirming a shared root cause despite different stall mechanisms.
Without LEO's cross-file dependency chain, this abstraction-layer bottleneck would be invisible in any single source file.

\begin{figure}[t]
\centering
\begin{subfigure}[t]{\columnwidth}
\begin{lstlisting}[language=C++,
  firstnumber=56,
  xleftmargin=15pt, framexleftmargin=0pt,
]
RAJA::kernel<ExecPolicy>(
  RAJA::make_tuple(range_Moments,
    range_Directions, range_Groups,
    range_Zones),
  [=] (IMoment nm, IDirection d,
       IGroup g, IZone z) {
(*@\colorbox{stallred!25}{~~phi(nm,g,z) += ell(nm,d)*psi(d,g,z);}@*)
  }
);
\end{lstlisting}
\caption{Kernel source code (\file{Kripke/Kernel/LTimes.cpp}).}
\label{fig:kripke-code}
\end{subfigure}

\vspace{6pt}\par

\begin{subfigure}[t]{\columnwidth}
\hspace{15pt}{\footnotesize\sffamily
\colorbox{usercolor!10}{\color{usercolor}\textbf{Kripke Code}}\quad
\colorbox{rajacolor!10}{\color{rajacolor}\textbf{RAJA Framework}}}\\[3pt]
{\footnotesize
\begin{tabular}{@{\hspace{15pt}}l@{\hspace{14pt}}l@{\hspace{14pt}}l@{}}
{\ttm\bfseries DFMA}                 & {\color{usercolor}\textit{LTimes.cpp:62}}          & {\color{stallred}\textbf{\qty{96.7}{\percent} stall cycles}}\\[1.5pt]
$\uparrow$~{\ttm LDG.E.64}          & {\color{usercolor}\textit{LTimes.cpp:62}}          & {\color{annotgray}global load (stalled)}\\[1.5pt]
$\uparrow$~{\ttm LEA.HI.X}          & {\color{rajacolor}\textit{TypedViewBase.hpp:216}}  & {\color{rajacolor}array index}\\[1.5pt]
$\uparrow$~{\ttm IADD3}             & {\color{rajacolor}\textit{Operators.hpp:369}}      & {\color{rajacolor}address offset}\\[1.5pt]
$\uparrow$~{\ttm IADD3}             & {\color{rajacolor}\textit{Iterators.hpp:291}}      & {\color{rajacolor}strided iteration}\\
\end{tabular}
}
\caption{LEO backward dependency slice.}
\label{fig:kripke-deps}
\end{subfigure}
\caption{LEO analysis of Kripke LTimes on NVIDIA GH200.
(a)~The highlighted line accounts for \qty{96.7}{\percent} of stall cycles.
(b)~LEO traces the stalling \instr{DFMA} instruction backward through a global load into three RAJA framework files (\textcolor{rajacolor}{purple}), revealing that address computation passes through RAJA's strided iterator.}
\label{fig:kripke-chain}
\vspace{-1em}
\end{figure}

\emph{Observation 4: Framework abstraction layers create bottlenecks invisible from source code alone.}
This pattern recurs across the case studies: QuickSilver's three-file pointer chain and HipKittens' compiler-lowered BF16 scalar loads both require cross-file backward tracing to diagnose.
LEO's ability to slice through abstraction boundaries is what makes these root causes actionable.

\section{Related Work}
\label{sec:related}

\paragraph{GPU Performance Tools}
Prior work covers adjacent pieces of LEO's design space, but not their combination. Vendor profilers such as Nsight Compute, ROCprofiler, and VTune report per-instruction stalls within a single ecosystem, but they do not trace those stalls backward to source-level causes or support GPUs from multiple vendors. GPA~\cite{zhou2021gpa} pioneered backward slicing for NVIDIA GPUs, while HPCToolkit's GPU measurement infrastructure~\cite{zhou2020hpctoolkit_gpu,zhou2021hpctoolkit_gpu_measurement,adhianto2024refining} provides LEO's profiling substrate. Cross-vendor performance studies~\cite{davis2025taking,kwack2025ai} document that the same application behaves differently across platforms, and DeepContext~\cite{zhao2025deepcontext} provides context-aware profiling for deep-learning workloads across platforms, but neither connects those differences to instruction-level root causes. PASTA~\cite{lin2026pasta} provides a modular analysis framework for accelerators but does not perform root-cause tracing. LEO occupies this missing intersection: a unified cross-vendor implementation that models vendor-specific PC-sampling semantics and performs backward slicing to actionable root causes.

\paragraph{Precise Event Sampling}
On CPUs, Sasongko et al.~\cite{sasongko2023} analyzed Intel PEBS and AMD IBS in terms of accuracy, stability, and overhead. We adopt a similar microbenchmark-driven perspective, but focus on GPU PC sampling facilities. Binary instrumentation frameworks such as SASSI~\cite{stephenson2015sassi}, NVBit~\cite{villa2019nvbit}, and Intel GTPin~\cite{gtpin2024,gorshkov2019gpu_hotspots} offer another route to fine-grained GPU measurement; however, such instrumentation-based tools are not well suited for measuring performance.

\paragraph{GPU Latency Hiding and Stall Analysis}
Volkov~\cite{volkov2016understanding} established the core theory of GPU latency hiding through massive multithreading and showed how Little's law distinguishes latency-bound from throughput-bound execution. Recent work on fine-grained stall accounting~\cite{cha2024gcstack,cha2025gcstack} addresses concurrent stall events that coarse mechanisms miss. GPUscout~\cite{sen2023gpuscout} uses CUPTI PC sampling to localize memory bottlenecks, DrGPUM~\cite{lin2023drgpum} guides memory optimization for GPU applications, and GhOST~\cite{chaturvedi2024ghost} proposes out-of-order warp scheduling to reduce stalls. Nayak and Basu~\cite{nayak2024oversynchronization} analyze unnecessary synchronization in GPU programs. LEO differs in aim: rather than characterizing stall behavior at a high level or redesigning scheduling, it traces stalls backward to the instructions that cause them.

\paragraph{Cross-Vendor Performance Portability}
Cross-vendor GPU performance has become increasingly important in the exascale era. Davis et al.~\cite{davis2025taking} evaluate seven programming models across NVIDIA and AMD GPUs, and Kwack et al.~\cite{kwack2025ai} benchmark 12 HPC and machine-learning applications across Frontier, Aurora, and Polaris. These studies establish the existence of cross-platform performance gaps. LEO complements such studies by explaining performance gaps at the instruction level.

\paragraph{Top-Down Analysis}
Yasin~\cite{yasin2014top} introduced top-down microarchitecture analysis for Intel CPUs, and Nowak et al.~\cite{nowak2015hierarchical} proposed hierarchical cycle accounting. For GPUs, Saiz et al.~\cite{saiz2022top} adapted top-down profiling to NVIDIA GPUs, and DrGPU~\cite{hao2023drgpu} extended the idea into a portable profiler. LEO is complementary rather than competitive: top-down methods classify where cycles go, whereas LEO traces specific stalls back to the instructions that cause them.

\paragraph{Program Slicing}
Weiser~\cite{weiser1984slicing} introduced program slicing as the computation of the program subset that affects a value at a given point. 
Horwitz et al.~\cite{horwitz1990interprocedural} extended slicing to interprocedural contexts. Cifuentes and Fraboulet~\cite{cifuentes1997intraprocedural} adapted slicing to binary executables, and Srinivasan and Reps~\cite{srinivasan2016improved} improved the precision of machine-code slicing. LEO adapts that line of work to GPU performance diagnosis by slicing backward from stalled instructions through register, predicate, and synchronization dependencies to root causes.\footnote{Some wording in this section was refined with assistance from Claude~\cite{claude_ai} and Gemini~\cite{gemini_ai}; the authors verified all references and the final text.}

\section{Conclusions}
\label{sec:conclusion}

This paper presents LEO, a cross-vendor GPU root-cause analyzer that uses backward slicing to identify the instructions responsible for stalls.
LEO leverages HPCToolkit's cross-vendor PC-sampling infrastructure to support NVIDIA, AMD, and Intel GPUs using a unified analysis layer. Across three platforms and 21 workloads, the evaluation shows that the same kernel can have different bottlenecks on different architectures and that LEO's diagnoses lead to actionable optimizations, with geometric-mean speedups of 1.73$\times$--1.82$\times$.

LEO builds dependency graphs from register dataflow, prunes them with opcode, barrier, latency, and execution constraints, and attributes blame with inverse-distance weighting. It also traces through vendor-specific synchronization mechanisms---AMD \instr{s\_waitcnt} counters, NVIDIA barrier bits, and Intel SWSB tokens---to reach memory operations that cause stalls.

LEO provides a unified tool for tracing instruction-level stalls back to actionable root causes on NVIDIA, AMD, and Intel GPUs. Our study exploring the utility of LEO's dependency chains to guide optimization using large language models is a secondary contribution. Notably, this study suggests that LEO's structured root-cause reports also can be useful to guide automated tuning. We therefore view LEO first as a diagnostic tool for developers and performance engineers, and only second as a foundation for automating performance optimization of GPU kernels.
In this sense, LEO's root-cause reports are best viewed as high-value diagnostic attributions that guide optimization, while formal causal validation of individual chains remains future work.

\textbf{Limitations.}
LEO has several limitations. 
\begin{itemize}
    \item It traces register dataflow, not memory dataflow, so pointer-chasing or indirect-memory root causes may be hidden.
    \item  Inverse-distance weighting is a heuristic; because LEO does not model branch probabilities, it can misattribute blame through branching control flow.
    \item LEO inherits the statistical limits of PC sampling: cold instructions may receive too few samples, and NVIDIA's CUPTI Activity API can distort concurrency by serializing kernel execution.
    \item GPU opcode classification depends on vendor-specific ISA tables that must evolve with new extensions.
    \item Our speedup evaluation uses optimizations designed by domain experts informed by LEO's analysis results. Future work should include controlled studies comparing optimization with and without LEO, and assessing the utility of LEO's feedback for novice developers.
    \item Ground truth for GPU stall causality is unavailable, so LEO does not quantify false positive rates directly.  Instead, we assess through (i)~case studies where LEO-guided optimizations yield substantial speedup, and (ii)~review of selected dependency chains by domain experts.
\end{itemize}

While our studies show that optimizations guided by LEO's dependency chains improve performance, this paper does not provide intervention-based validation for every reported chain. For that reason, reported root causes should  be interpreted as actionable attributions supported by sampling and dependence analysis rather than formal causal proofs.

Finally, our study exploring improvements to kernels by LLMs  guided by dependency chains from LEO employs only a single model: Gemini 3.1 Pro. A thorough evaluation of how well various large language models tune code when presented with dependency chains from LEO remains future work.

In the future, we plan to extend LEO's analysis with indirect-memory tracking, predicate-aware weighting, and temporal characterization of multi-GPU workloads.

\section*{Acknowledgment}
Large language model tools were used in a limited way to suggest wording revisions for selected prose passages, including parts of the Related Work section.
The authors wrote the initial draft, verified the cited literature, and made all final decisions about wording and content.
No AI tool was used to generate the scientific ideas, methodology, experiments, figures, tables, results, or conclusions.

\bibliographystyle{IEEEtran}
\balance
\bibliography{refs}

\end{document}